\documentclass[aps,pre,preprint,groupedaddress]{revtex4-1}

\usepackage{graphics}
\usepackage{graphicx}

\begin{document}


\title{A test of multiple correlation temporal window characteristic of non-Markov processes}


\author{F.T.Arecchi}
\email[]{tito.arecchi@ino.it}
\altaffiliation{Dipartimento di Fisica Universit\`a di Firenze}
\affiliation{INO-CNR Largo E.Fermi 6 50125 Firenze}

\author{A.Farini}
\email[]{alessandro.farini@ino.it}
\affiliation{INO-CNR Largo E.Fermi 6 50125 Firenze}

\author{N.Megna}
\email[]{nicola.megna@ino.it}
\affiliation{INO-CNR Largo E.Fermi 6 50125 Firenze}


\date{\today}

\begin{abstract}
We introduce a sensitive test of memory effects in successive events. 
The test consists of a combination $K$ of binary correlations at successive times. 
$K$ decays monotonically from $K=1$  for uncorrelated events as a Markov process; 
whereas memory effects provide a temporal window with  $K>1$. 
For a monotonic memory fading, $K<1$ always. 
Here we report evidence of a $K>1$ temporal window in cognitive tasks consisting of the visual identification of the front face of the Necker cube after a previous presentation of the same.
The $K>1$ behaviour is maximal at an inter-measurement time $\tau$ around 2 sec with inter-subject differences. 
The $K>1$ persists over a time window of 1 sec around $\tau$; outside this window the $K<1$ behaviour is recovered. 
The universal occurrence of a $K>1$ window in pairs of successive perceptions suggests that, at variance  with single visual stimuli eliciting a suitable response, a pair of stimuli  shortly separated in time displays mutual correlations.
\end{abstract}


\maketitle

\section{Introduction}
The $K$-test highlights memory effects in successive events. 
It consists of a combination of binary correlations at successive times. 
$K$ decays monotonically from $K=1$  for uncorrelated events as a Markov process; whereas memory effects provide a transient peak above $K=1$; the peak duration measures the time extent of the memory effect.
 
We apply the $K$ test to successive perceptions showing that,at variance with isolated presentations, sequential presentations display mutual correlations over a characteristic time interval.

\section{A test that discriminates sequences of uncorrelated perceptions from sequences of pairwise correlated perceptions}
\begin{figure}	
	\includegraphics[width=\linewidth]{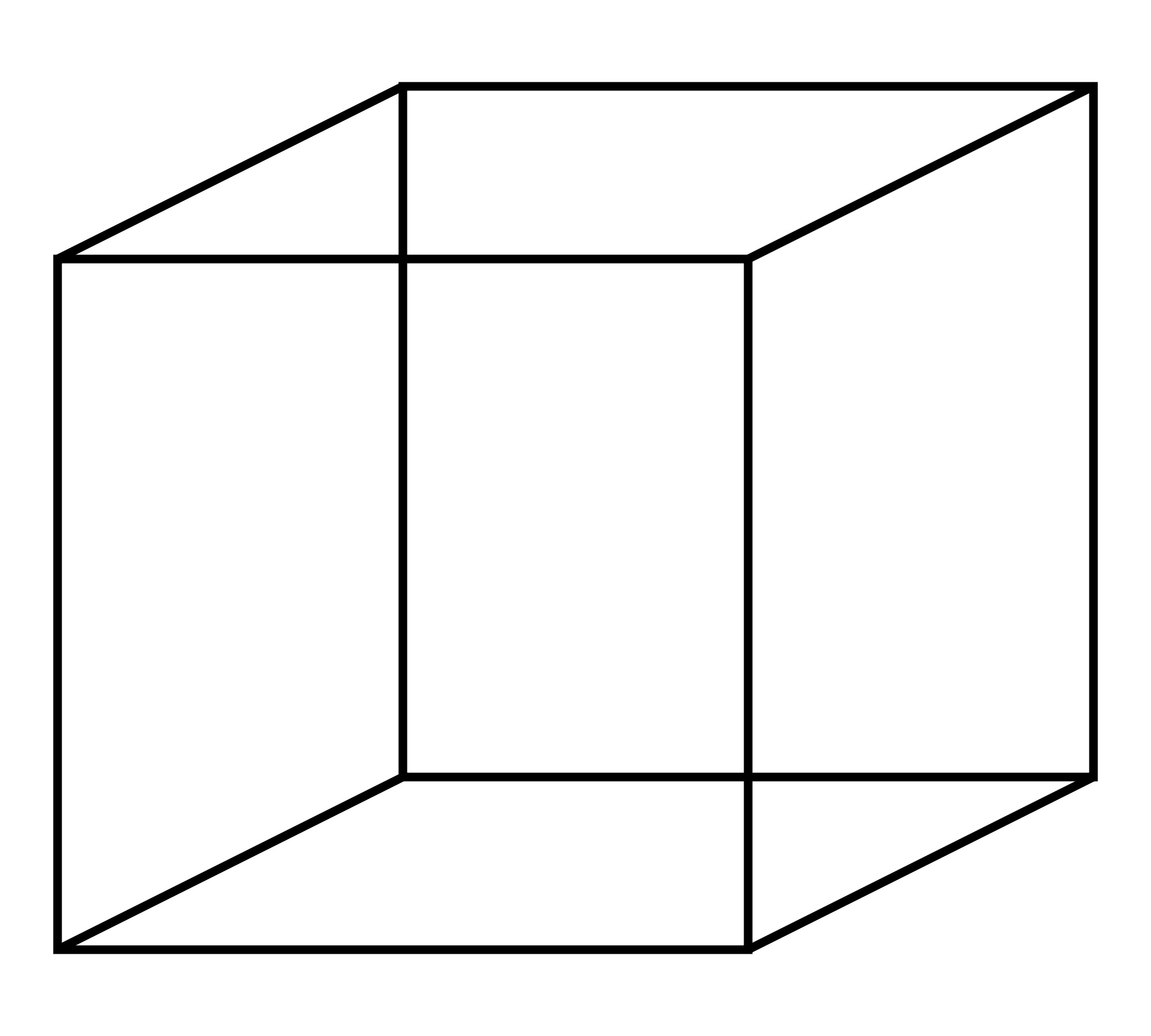}
	\caption{The Necker Cube}
	\label{fig:neckercube}
\end{figure}

We want to obtain, from the perception of a visual stimulus, a sequence of binary values $(+1,-1)$. 
The simplest way is to submit to subjects' observation a bistable figure\cite{long2004enduring}. 
Our choice was the Necker cube\cite{necker1832}, where the perceived perspective of the front face of the cube (Fig.\ref{fig:neckercube}) alternates between two different options as either facing upward and to the right or downward and to the left: 
we attribute respectively measurement values $Q=\pm 1$.

In our experiment the measurement corresponds to obtain from a subject an answer about which face he/she is viewing as front face at a given time. 
We aim at providing a sensitive test of correlations between successive presentations of the cube. 
The key here is that we have measurements at two different times, and one or more times between the first and last measurement. 
Let us perform measurement at three successive times $t_1<t_2<t_3$ with ISI (inter-stimulus interval) $=t_2-t_1=t_3-t_2$ 
(the relevance for the experiment of these intervals will be explained in the next section). 
For instance, the correlation $C_{13}$ between times $t_{1}$  and $t_{3}$ for $N$ realizations of the experiment, reads
\begin{equation}
C_{13}=\frac{1}{N}\sum\limits_{r=1}^N Q_{r}(t_1) Q_{r}(t_3)
\end{equation}
We introduce the combination $K$ as $K=C_{12}+C_{23}-C_{13}$. 
In terms of actual measurements $Q$'s, $K$ is given by
\begin{equation}
K=\frac{1}{N}\sum\limits_{r=0}^N (Q(t_1) Q(t_2)+Q(t_2) Q(t_3)-Q(t_1) Q(t_3))
\label{eq:K}
\end{equation}
The $K$-test is the most elementary test to be performed on sequences of binary data. 
In a different context, namely in a temporal version of a quantum Bell test, 
it  was introduced by Leggett and Garg\cite{leggett1985} and exploited by many groups\cite{emary2014leggett}. 
However in the Leggett Garg case each term of the three sums which yield $C_{12}$, $C_{23}$, and $C_{13}$ is measured  sequentially in a single measurement and the ensemble average is carried over the the whole measurement. 
Instead we are carrying separate averages over the three components $C_{ij}$, so that the resemblance with the Leggett-Garg test is purely formal.

\begin{figure}
	\includegraphics{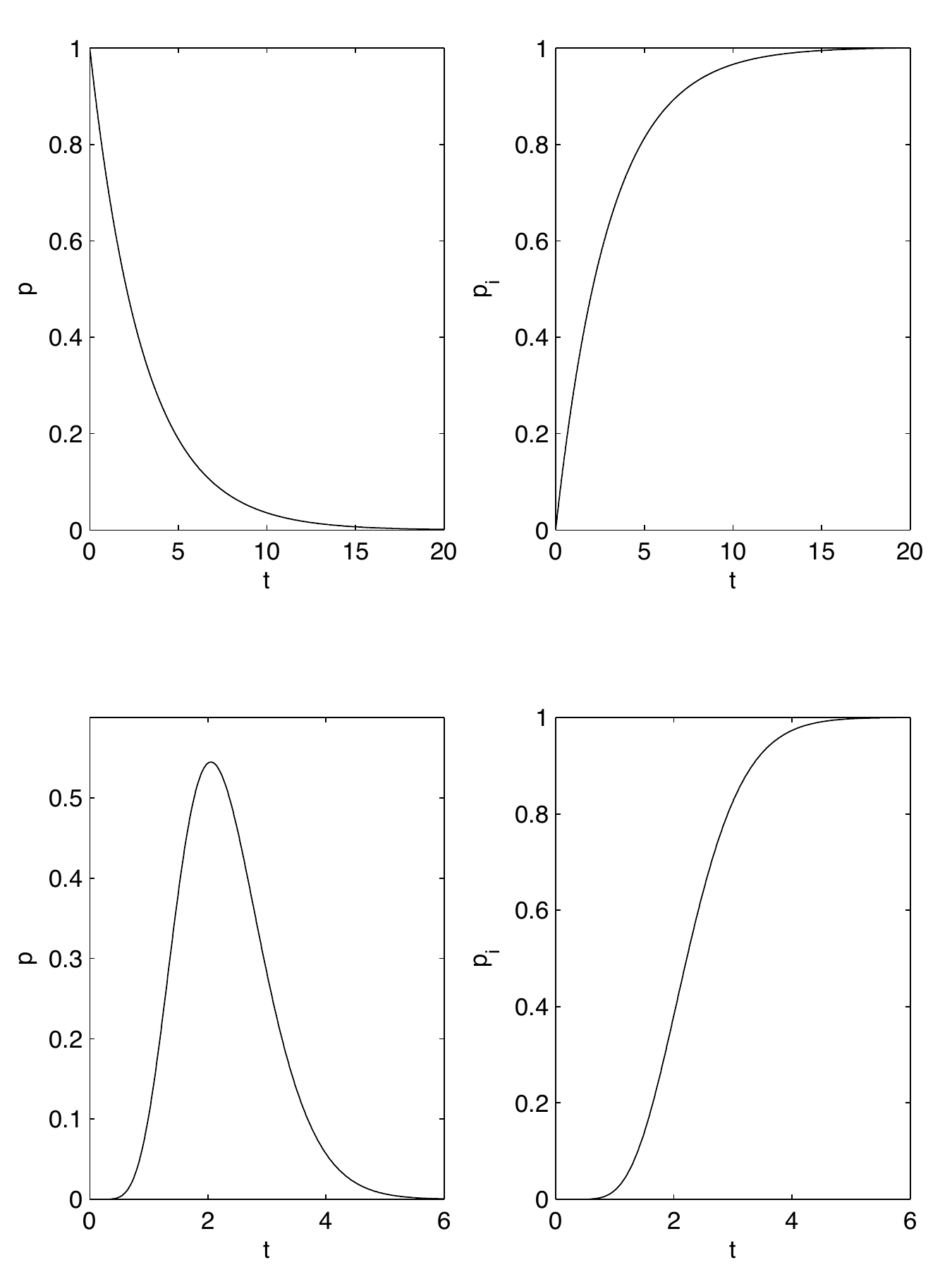}
	\caption{a) (up-left) Probability $p$ to have a single switch at a time $t$ for a sequence of random switches (uniform probability per unit time) b) (up-right) Probability $p_i$that at least a switch is occurred at a time $t$ for a sequence of random switches (uniform probability per unit time). The function corresponds to the integral of function in a) c)(down-left) Probability $p$ to have a single switch at a time $t$ for the gamma distribution. d) (down-right) Probability $p_i$ that at least a switch is occurred at a time $t$ for the gamma distribution. The function corresponds to the integral of function in c)}
	\label{fig:pdfcdf}
\end{figure}

In order to understand what we can expect from our visual experiment, we analyze some numerical simulations obtained starting from some models related to Necker cube. 
A possible model is that the switch from +1 to -1 occurs randomly with a uniform distribution in time (Poisson process). 
In this condition the probability $p(t)$ of the first switch at a given time $t$ decays as a $e^{\frac{-t}{\tau}}$ where $\tau$ is the average occurrence of switches (Fig.\ref{fig:pdfcdf}a). 
The probability $p_i(t)$ that at least a switch is occurred at a time $t$ is given by the integral of the previous function which asymptotically yields 1 (Fig.\ref{fig:pdfcdf}b).
In the case of human subjects viewing the Necker cube a well verified model\cite{borsellino1972,gigante2009} is that the probability of the first switch $p$ is given by the gamma function plotted in Fig.\ref{fig:pdfcdf}c and the integrated probability $p_i$ is given in Fig.\ref{fig:pdfcdf}d. 
At variance with Fig.\ref{fig:pdfcdf}b, Fig.\ref{fig:pdfcdf}d displays an initial correlation with the $t=0$ event, followed by a sharp increase. 
While Fig.\ref{fig:pdfcdf}b  is uniformly convex, the presence of memory effects is associated with the change of curvature (from concave to convex) of Fig.\ref{fig:pdfcdf}d.
In collecting data on real subjects the differences between Fig.\ref{fig:pdfcdf}b and Fig.\ref{fig:pdfcdf}d may be masked by observation features. 
Furthermore, the gamma function of Ref\cite{borsellino1972,gigante2009} corresponds to a continuous presentation of the Necker cube. 
Instead, as we explain later, we are interested to a flashing presentation, which seems more appropriate to mimic the variety of perceptions during an observation. 
We thus look for a suitable combination of correlation functions which provides a sensitive test not only for discriminating upper from lower probabilities in Fig.\ref{fig:pdfcdf} but also for discriminating between continuous and flashing presentation of the Necker cube. 

\begin{figure}	
		\includegraphics[width=\linewidth]{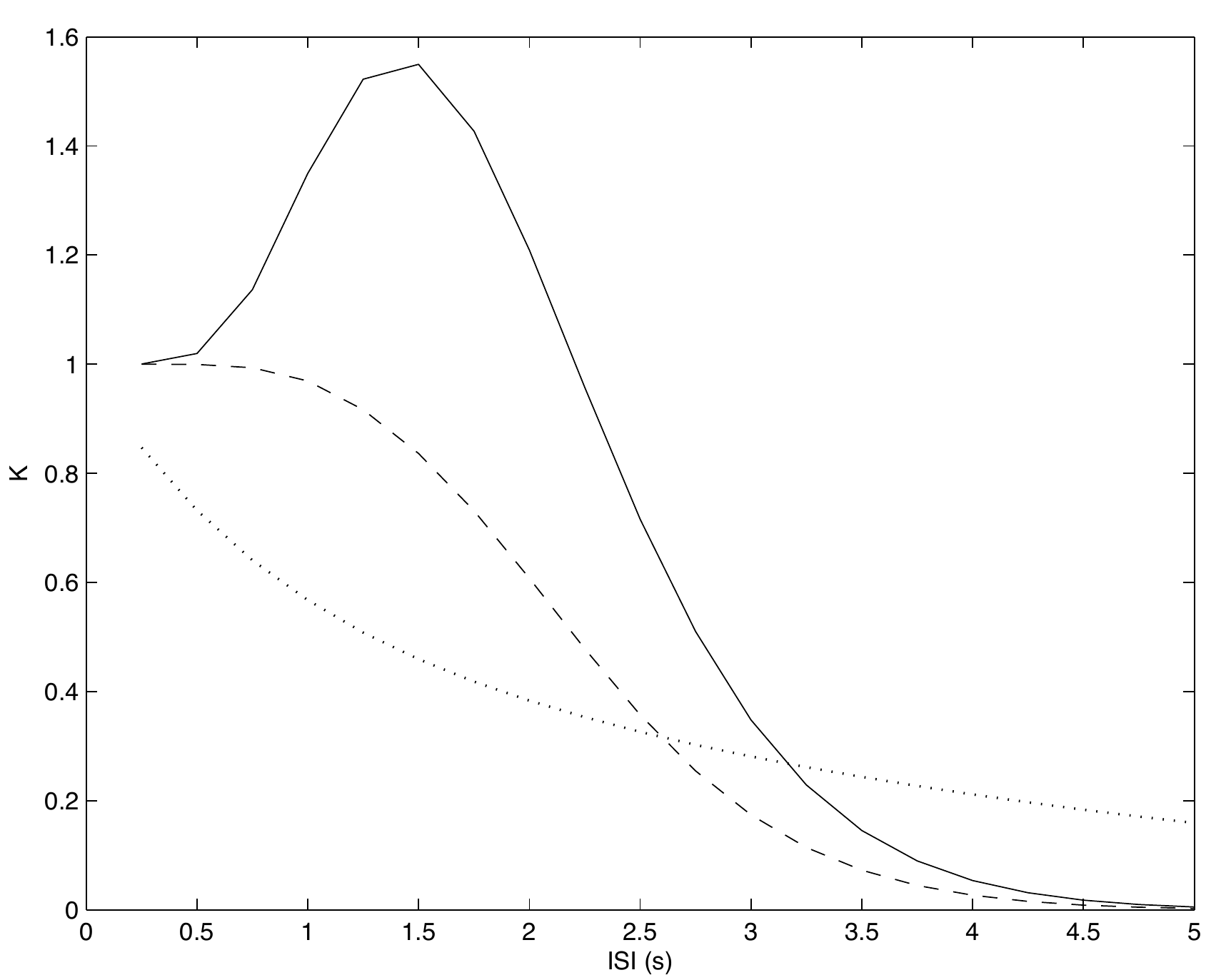}
		\caption{a) Computational $K$ for random switches (dotted line),  b) for gamma distribution in a continuous presentation (dashed line)  and c) for a gamma distribution in a flashing presentation (continuous line)}
		\label{fig:kvari}
\end{figure}

We correlate the data at three different times equally spaced $t_1$ (corresponding to the starting time) $t_2$ and $t_3$, 
building the corresponding correlation functions. 
If we set $Q(t_1)=1$, $C_{12}=1\cdot(1-p_i(t_2-t_1))$. 
Notice that since $t_2-t_1$ equals $t_3-t_2$, a continuous presentation of a Necker cube yields $C_{23}=(1-p_i(t_2-t_1))(1-p_i(t_3-t_2))$ whereas if we present the Necker cube in a flashing way, we have a reset in time and probability and hence $C_{12}=C_{23}$. 
From these assumptions we can compute $K$ using Eq.\ref{eq:K}. 
It results for the memory-less case $K<1$ always (Fig.\ref{fig:kvari}a), 
for a continuous presentation of the Necker cube again $K<1$ (Fig.\ref{fig:kvari}b), 
instead for the flashing presentation $K$ goes above 1 within a temporal window (Fig.\ref{fig:kvari}c). 
Thus the $K$-test appears as a very sensitive one for extracting memory effects.

Let us discuss the relevance of $K$.
In cognitive perceptions,the associated motor response emerges from a Bayesian inference\cite{arecchi2007physics,griffiths2008} 
whereby the conditional probability that a datum $d$ follows from an hypothesis $h$ is a memorized algorithm. 
Such a process requires an elaboration lasting a few hundred milliseconds, up to about 1 second\cite{lachaux1999}.  
On the contrary, in the exposure to a correlated sequences of perception, 
the interpretation of a piece is not done in terms of an already established algorithm but it depends on the previous piece. 
The build up of such a specific correlation has been called ``inverse Bayes inference''\cite{arecchi2010}. 

We aim at building a sensitive test of a memory binding lasting for such a characteristic time in the exposure to a bistable image. 

We consider an uncorrelated sequence of perceptions as the standard cognitive state of a brainy animal, 
each perception being built by combining a bottom-up sensorial stimulus with a top-down instruction provided by the long term memory. 
A cognitive task consists of a chain of perceptions, each one eliciting a motor reaction\cite{rodriguez1999perception}.
Successive perceptions keep a Bayesian structure; a sequence of them is then Markovian, 
in the sense that a previous perception does not affect in a structural way the next one \cite{griffiths2008,kording2006}.

On the contrary, when successive perceptions display correlations, 
a bridge has been provided by the short term memory which has retrieved the previous perception. 
We hypothesize that this short term bridge is the basis of language, 
so that referring to correlated sequences of $+/-1$ states, we are in presence of the most elementary language, 
consisting of binary words\cite{poppel2004lost}; 
whenever a piece of a linguistic text  is compared with a previous piece retrieved by the short term memory in order to infer a comparison a judgment emerges.

Perception and judgments have been associated, respectively, with a Bayes inference and with an inverse Bayes inference \cite{arecchi2010,arecchi2011phenomenology}. 
These last references discuss also the two associated times, namely, around 1 sec for perception \cite{koch2004,dehaene2003,lachaux1999} and around 3 sec for judgments. 
The 3 sec intervals in the judgment operations had been suggested by the extensive analyses of this cognitive time provided by Ernst Poeppel \cite{poppel2004lost,poppel1997hierarchical,poppel2009pre} .

To avoid semantic biases, we take as ``linguistic text'' an ordered sequence of binary signals (sequential presentations of the Necker cube). In Sec.\ref{materials} and \ref{results} we describe the methods and the results. In Sec.\ref{discussion} we provide cues for the temporal window of $K>1$ and for the approximate location of its peak value. Finally we discuss possible relations between the peak time of the $K>1$ curve and the peak in the statistics of pauses in linguistic endeavours.

\section{Materials and methods} \label{materials}
\begin{figure}
	\includegraphics[width=\linewidth]{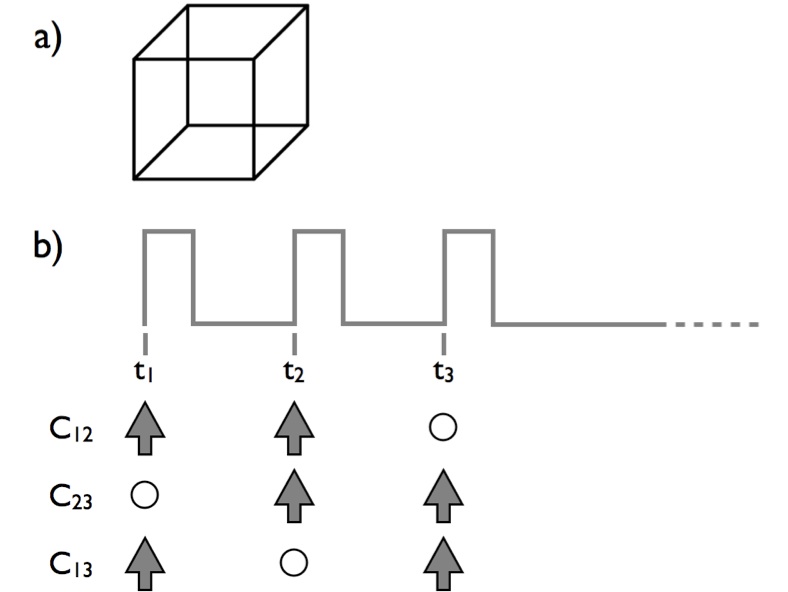}
	\caption{a) The Necker Cube b) Experimental procedure. Sequence of three successive presentations of the Necker cube (represented as square pulses, lasting 0.30 sec each)
	separated by time intervals
	$ISI=t_{2}-t_{1}=t_{3}-t_{2}$,
	adjustable from 1 sec up. The vertical arrows denote a sharp acoustic signal acting as a stimulus that demands the subject to press either button corresponding to
	the perceived front face of the cube. The circles denote the
	presentation of the cube in the absence of the acoustic signal. The
	three sequences correspond to $C_{12}$, $C_{23}$
	and $C_{13}$ respectively. The sequences are repeated after
	a time much longer than $t_{2}-t_{1}$.}
	\label{fig:time}
\end{figure}

The stimulus presented to the observers was the Necker cube (Fig.\ref{fig:time}a) whereby the perceived  perspective of the front face of the cube alternates between two different options as either facing  upward and to the right or downward and to the left. 
These two percepts alternate in time. 
All visual  stimuli were displayed by a 21 inches colour CRT monitor (Sony GDM-F500 800*600 pixels, refresh rate 100 Hz) and generated by a Personal Computer. 
The experiment was written using the Psychophysics Toolbox 2.54 extensions \cite{brainard1997,pelli1997,kleiner2007}. 

The Necker cube was defined by black lines and covered an area of $5^{\circ}\times 5^{\circ}$. 
Observers watched binocularly the display at a viewing distance of 57 cm. 
The stimulus was displayed at the center of the screen on a mid gray background with luminance of $25 cd/m^2$. 
No fixation point was presented, as observers were asked to passively inspect the stimulus on the screen, with natural eye movements.
Subjects viewed for 120 times (N) per condition a sequence of three Necker cubes at three different times ($t_1$, $t_2$, $t_3$). 
Each presentation of the cube had a duration of 0.3 seconds and was separated from the next presentation by a variable interstimulus interval (ISI) with the condition that $t_2-t_1=t_3-t_2$. 
Each sequence was followed by a blank interval (no stimulus displayed) of 5 seconds, well above the maximal ISI.
For two of the three stimuli of the sequence, an acoustic beep was presented to the subjects (Fig.\ref{fig:time}b), 
asking which of the two different percept they had for the corresponding stimulus, 
whence pressing a key on the keyboard if they perceived a cube with an upward-right front face or a different key if they perceived a downward-left front face.  
The subjects, for a total number of 21,  were all students of the University of Firenze, 
with normal or corrected to normal vision, without pre-information of the aim of the study. 
Each subject gave a written informed consent to participate, and the experiment was done according to the declaration of Helsinki. 
The ethics committee of the National Institute of Optics specifically approved this study. 
Incidentally, 21 subject could appear a small number, but in visual psychophysics some important effects were shown using only two subjects 
(e.g.\cite{neri1998seeing}) also the CIE standard observers (used to calculate the efficiency of lamps) are averages based on experiments with small numbers (around 15-20) of people\cite{wyszecki1982color}.

\begin{figure}
	\includegraphics[width=\linewidth]{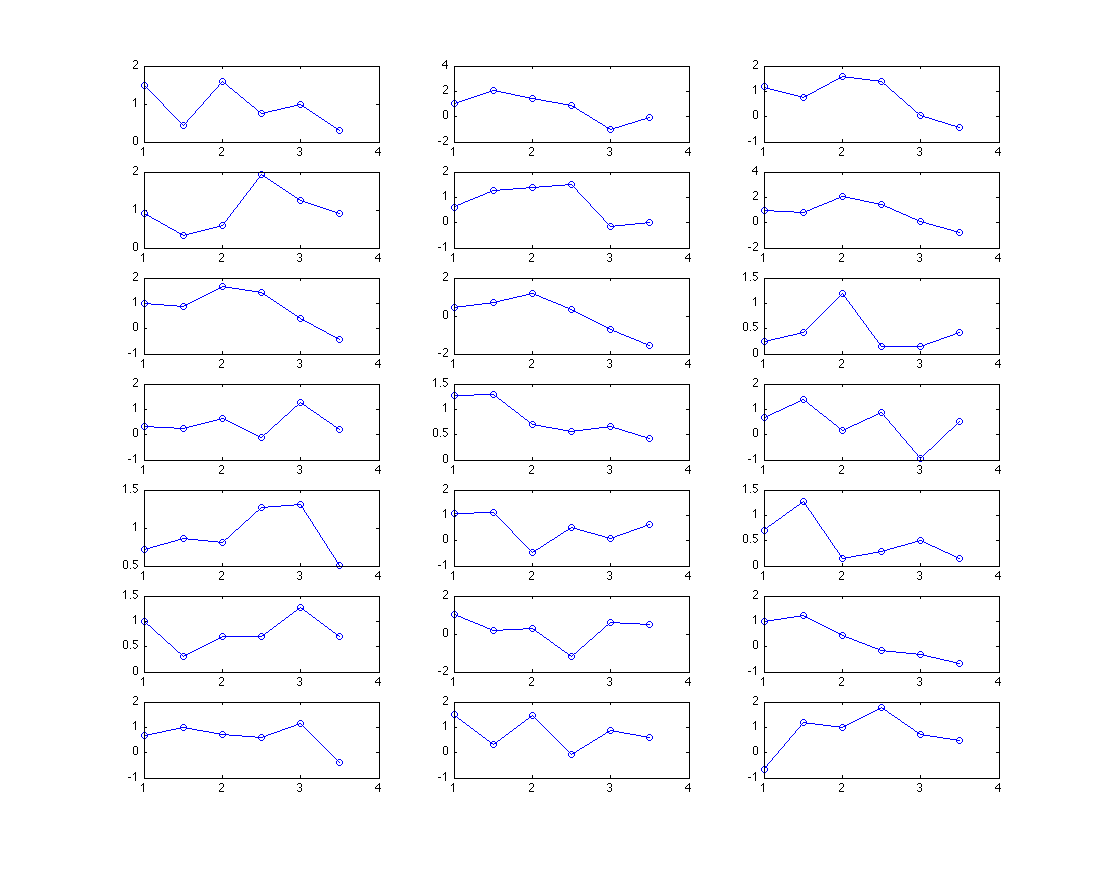}
	\caption{$K$ values corresponding to the six selected interstimulus intervals for
	each of the 21 subjects that participated to the experiment. 
	The correlation $C_{12}$ results as a sum of products $Q1*Q2$, similarly for $C_{23}$  and $C_{13}$ . By these data introduced into Eqs. (1) and (2), we evaluate $K$ for each ISI.}
	\label{fig:kvalues}
\end{figure}

\section{Results} \label{results}
We expose 21 normal subjects to a balanced presentation
of a Necker cube \cite{necker1832,farini16}, a wire frame cube with no depth cues, and request to press two different buttons,
depending on the cube face perceived as the front one, anytime a short
audio signal alerts the subject (Fig.\ref{fig:time}).

In Fig.\ref{fig:kvalues} we report $K$ versus the inter-stimulus interval (ISI) $t_{2}-t_{1}=t_{3}-t_{2}$ for 21 subjects.  
For each subject, successive presentations of the sequence of Fig.\ref{fig:time} are separated by a blank time of 5 sec, well above the maximum ISI
explored. 
In Fig.\ref{fig:time} the time duration of each cube presentation is 0.30 sec. 
One might guess that the timing is ruled  by the acoustic warning signals (arrows of Fig.\ref{fig:time}) and the flashed presentation of the cube does not matter. 
In fact, we have tested  also with a continuous presentation of  the cube; however, in such a case all subjects yield $K$ values consistently below 1 and located between 1.0 and -1.5, 
without any correlation with the ISI interval, in accordance with the expectation of Fig.\ref{fig:kvari}b.

The  $K<1$ result  for the continuous presentation of the Necker cube reinforces the guess that the search in the semantic 
space has to be performed on different items. We have obtained the same result ($K<1$) testing the subjects with another bistable figure, 
the rotating sphere \cite{Leopold:2002stable}, with continuous presentation. On the other hand a flashed presentation of the rotating sphere as 
done so far with appropriate software implies switching times much longer than the times here considered and thus it loses any linguistic relevance.

\begin{figure}
	\includegraphics[width=\linewidth]{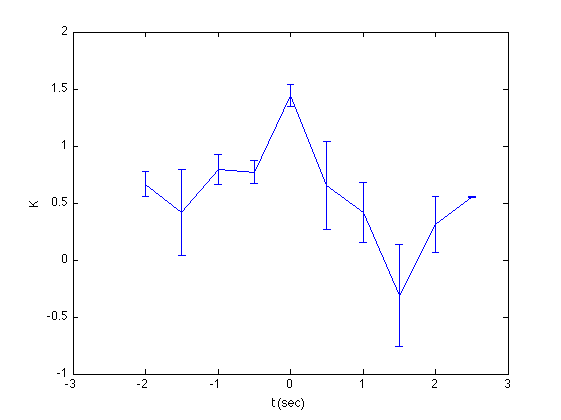}
	\caption{ $K$ values in correspondence of the six tested interstimulus intervals. These values have been obtained by pooling data from the 21 subjects into a unique meta-observer.
	The data are also pooled around the peak time, here denoted as $t=0$.  The error bars are the variance. $K>1$ is statistically significant at the peak.}
	\label{fig:kvaluesmeta}
\end{figure}

\begin{figure}
	\includegraphics[width=\linewidth]{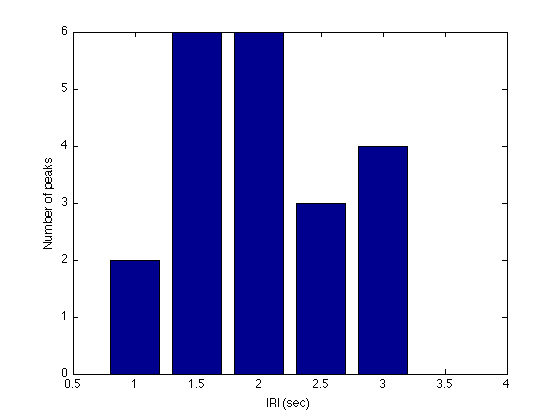}
	\caption{Distributions $\tau$ of the maxima of $K$ (Fig.\ref{fig:time}).}
	\label{fig:peaks}	
\end{figure}

Due to the fact that every subject could have a different decorrelation time we have created a meta-subject response aligning the peaks at a time $t=0$ and plotting
the averaged $Ks$ for negative and positive time separations from the
peak $\tau$ .  The $K>1$ result is statistically significant at the peak(Fig.\ref{fig:kvaluesmeta}).  
In Fig \ref{fig:peaks} we plot the statistical distribution of  $\tau$ for the 21 subjects.

The mean switching time, related to $C_{12}$, has been previously measured \cite{borsellino1972,leopold1999}. As for the single 
time correlation $C_{12}$, a deviation from a purely classical statistics, called Quantum Zeno effect, has 
been reported \cite{atmanspacher2004}. It refers however to time scales below 1 sec, well within the perception regime, thus it has no relevance for linguistic purposes.

\section{Discussion: Why $\tau$, a qualitative explanation}  \label{discussion}
We elaborate on the $K>1$ effect around 2 sec. 
A very striking new fact is that the peak $\tau$ is close to the maximal value of the statistical distribution of pauses in several linguistic endeavours\cite{poppel1997hierarchical}.

From the neuroscientific side, no attempt has been made to tackle the sharp discrepancy between perception times, for which several NCC have been devised \cite{koch2004} and the judgment times necessary to build meanings out of sequences of consecutive linguistic pieces.

Let us speculate about the time location $\tau$. 
Since $\tau$ corresponds to the maximal perturbation that a word exerts on a later word, then the human linguistic processing is organized so that $\tau$
is the most appropriate coupling time between two successive groups of words. 
A time window of about 1 sec surrounds $\tau$; it represents the interval over which $K>1$. 
Thus $K>1$ is in fact a short term memory indicator. 
Our cognitive abilities seem geared in such a way that two adjacent groups of words within that window interact in a non-Bayesian way \cite{arecchi2011phenomenology}. 
Away from that window, that is, for shorter and longer separations of word groups, a classical Bayesian strategy \cite{griffiths2008,kording2006} applies. 
We hypothesize that, in order to make sense of a text, two successive pieces must be separated by a time of the order of $\tau$.

The individual lumps  generate timeless
perceptions by themselves and the time separations from one to the next
are the relevant ones in the $K$ test here discussed.

Thus the $K > 1$ occurrence is a strategy to correlate successive lumps - as the sequential pieces of a linguistic text- one another, overcoming (or, better to say, re-adjusting) the arousal that each piece provokes by itself. Our cognitive abilities seem to be geared for two different tasks, namely,
\begin{enumerate}
	\item	within the single lump, make the best use of the incoming sensorial information exploiting previous information stored in the long term memory (perception), and 
	\item	combine the current perception with the correlation to   the previous lump of the  linguistic sequence (short term memory bridge); the two lumps may interact with different weights depending on the semantic content of the sequence we are exposed to. 
\end{enumerate}

The $\tau$ measured by us in a sequence with no semantic content provides the first experimental evidence of mechanism 2.

\bibliography{bibliolgi}

\begin{thebibliography}{27}%
\makeatletter
\providecommand \@ifxundefined [1]{%
 \@ifx{#1\undefined}
}%
\providecommand \@ifnum [1]{%
 \ifnum #1\expandafter \@firstoftwo
 \else \expandafter \@secondoftwo
 \fi
}%
\providecommand \@ifx [1]{%
 \ifx #1\expandafter \@firstoftwo
 \else \expandafter \@secondoftwo
 \fi
}%
\providecommand \natexlab [1]{#1}%
\providecommand \enquote  [1]{``#1''}%
\providecommand \bibnamefont  [1]{#1}%
\providecommand \bibfnamefont [1]{#1}%
\providecommand \citenamefont [1]{#1}%
\providecommand \href@noop [0]{\@secondoftwo}%
\providecommand \href [0]{\begingroup \@sanitize@url \@href}%
\providecommand \@href[1]{\@@startlink{#1}\@@href}%
\providecommand \@@href[1]{\endgroup#1\@@endlink}%
\providecommand \@sanitize@url [0]{\catcode `\\12\catcode `\$12\catcode
  `\&12\catcode `\#12\catcode `\^12\catcode `\_12\catcode `\%12\relax}%
\providecommand \@@startlink[1]{}%
\providecommand \@@endlink[0]{}%
\providecommand \url  [0]{\begingroup\@sanitize@url \@url }%
\providecommand \@url [1]{\endgroup\@href {#1}{\urlprefix }}%
\providecommand \urlprefix  [0]{URL }%
\providecommand \Eprint [0]{\href }%
\providecommand \doibase [0]{http://dx.doi.org/}%
\providecommand \selectlanguage [0]{\@gobble}%
\providecommand \bibinfo  [0]{\@secondoftwo}%
\providecommand \bibfield  [0]{\@secondoftwo}%
\providecommand \translation [1]{[#1]}%
\providecommand \BibitemOpen [0]{}%
\providecommand \bibitemStop [0]{}%
\providecommand \bibitemNoStop [0]{.\EOS\space}%
\providecommand \EOS [0]{\spacefactor3000\relax}%
\providecommand \BibitemShut  [1]{\csname bibitem#1\endcsname}%
\let\auto@bib@innerbib\@empty
\bibitem [{\citenamefont {Long}\ and\ \citenamefont
  {Toppino}(2004)}]{long2004enduring}%
  \BibitemOpen
  \bibfield  {author} {\bibinfo {author} {\bibfnamefont {G.~M.}\ \bibnamefont
  {Long}}\ and\ \bibinfo {author} {\bibfnamefont {T.~C.}\ \bibnamefont
  {Toppino}},\ }\href@noop {} {\bibfield  {journal} {\bibinfo  {journal}
  {Psychological bulletin}\ }\textbf {\bibinfo {volume} {130}},\ \bibinfo
  {pages} {748} (\bibinfo {year} {2004})}\BibitemShut {NoStop}%
\bibitem [{\citenamefont {Necker}(1832)}]{necker1832}%
  \BibitemOpen
  \bibfield  {author} {\bibinfo {author} {\bibfnamefont {L.}~\bibnamefont
  {Necker}},\ }\href@noop {} {\bibfield  {journal} {\bibinfo  {journal} {The
  London and Edinburgh Philosophical Magazine and Journal of Science}\ }\textbf
  {\bibinfo {volume} {1}},\ \bibinfo {pages} {329} (\bibinfo {year}
  {1832})}\BibitemShut {NoStop}%
\bibitem [{\citenamefont {Leggett}\ and\ \citenamefont
  {Garg}(1985)}]{leggett1985}%
  \BibitemOpen
  \bibfield  {author} {\bibinfo {author} {\bibfnamefont {A.}~\bibnamefont
  {Leggett}}\ and\ \bibinfo {author} {\bibfnamefont {A.}~\bibnamefont {Garg}},\
  }\href@noop {} {\bibfield  {journal} {\bibinfo  {journal} {Physical Review
  Letters}\ }\textbf {\bibinfo {volume} {54}},\ \bibinfo {pages} {857}
  (\bibinfo {year} {1985})}\BibitemShut {NoStop}%
\bibitem [{\citenamefont {Emary}\ \emph {et~al.}(2014)\citenamefont {Emary},
  \citenamefont {Lambert},\ and\ \citenamefont {Nori}}]{emary2014leggett}%
  \BibitemOpen
  \bibfield  {author} {\bibinfo {author} {\bibfnamefont {C.}~\bibnamefont
  {Emary}}, \bibinfo {author} {\bibfnamefont {N.}~\bibnamefont {Lambert}}, \
  and\ \bibinfo {author} {\bibfnamefont {F.}~\bibnamefont {Nori}},\ }\href@noop
  {} {\bibfield  {journal} {\bibinfo  {journal} {Reports on Progress in
  Physics}\ }\textbf {\bibinfo {volume} {77}},\ \bibinfo {pages} {016001}
  (\bibinfo {year} {2014})}\BibitemShut {NoStop}%
\bibitem [{\citenamefont {Borsellino}\ \emph {et~al.}(1972)\citenamefont
  {Borsellino}, \citenamefont {Marco}, \citenamefont {Allazetta}, \citenamefont
  {Rinesi},\ and\ \citenamefont {Bartolini}}]{borsellino1972}%
  \BibitemOpen
  \bibfield  {author} {\bibinfo {author} {\bibfnamefont {A.}~\bibnamefont
  {Borsellino}}, \bibinfo {author} {\bibfnamefont {A.}~\bibnamefont {Marco}},
  \bibinfo {author} {\bibfnamefont {A.}~\bibnamefont {Allazetta}}, \bibinfo
  {author} {\bibfnamefont {S.}~\bibnamefont {Rinesi}}, \ and\ \bibinfo {author}
  {\bibfnamefont {B.}~\bibnamefont {Bartolini}},\ }\href@noop {} {\bibfield
  {journal} {\bibinfo  {journal} {Biological Cybernetics}\ }\textbf {\bibinfo
  {volume} {10}},\ \bibinfo {pages} {139} (\bibinfo {year} {1972})}\BibitemShut
  {NoStop}%
\bibitem [{\citenamefont {Gigante}\ \emph {et~al.}(2009)\citenamefont
  {Gigante}, \citenamefont {Mattia}, \citenamefont {Braun},\ and\ \citenamefont
  {Del~Giudice}}]{gigante2009}%
  \BibitemOpen
  \bibfield  {author} {\bibinfo {author} {\bibfnamefont {G.}~\bibnamefont
  {Gigante}}, \bibinfo {author} {\bibfnamefont {M.}~\bibnamefont {Mattia}},
  \bibinfo {author} {\bibfnamefont {J.}~\bibnamefont {Braun}}, \ and\ \bibinfo
  {author} {\bibfnamefont {P.}~\bibnamefont {Del~Giudice}},\ }\href {\doibase
  10.1371/journal.pcbi.1000430} {\bibfield  {journal} {\bibinfo  {journal}
  {PLoS Comput Biol}\ }\textbf {\bibinfo {volume} {5}},\ \bibinfo {pages}
  {e1000430} (\bibinfo {year} {2009})}\BibitemShut {NoStop}%
\bibitem [{\citenamefont {Arecchi}(2007)}]{arecchi2007physics}%
  \BibitemOpen
  \bibfield  {author} {\bibinfo {author} {\bibfnamefont {F.}~\bibnamefont
  {Arecchi}},\ }\href@noop {} {\bibfield  {journal} {\bibinfo  {journal} {The
  European Physical Journal Special Topics}\ }\textbf {\bibinfo {volume}
  {146}},\ \bibinfo {pages} {205} (\bibinfo {year} {2007})}\BibitemShut
  {NoStop}%
\bibitem [{\citenamefont {Griffiths}\ \emph {et~al.}(2008)\citenamefont
  {Griffiths}, \citenamefont {Kemp},\ and\ \citenamefont
  {Tenenbaum}}]{griffiths2008}%
  \BibitemOpen
  \bibfield  {author} {\bibinfo {author} {\bibfnamefont {T.}~\bibnamefont
  {Griffiths}}, \bibinfo {author} {\bibfnamefont {C.}~\bibnamefont {Kemp}}, \
  and\ \bibinfo {author} {\bibfnamefont {J.}~\bibnamefont {Tenenbaum}},\
  }\enquote {\bibinfo {title} {Bayesian models of cognition},}\ \ (\bibinfo
  {publisher} {Cambridge Univ Press},\ \bibinfo {year} {2008})\ pp.\ \bibinfo
  {pages} {59--100}\BibitemShut {NoStop}%
\bibitem [{\citenamefont {Lachaux}\ \emph {et~al.}(1999)\citenamefont
  {Lachaux}, \citenamefont {Rodriguez}, \citenamefont {Martinerie},
  \citenamefont {Varela} \emph {et~al.}}]{lachaux1999}%
  \BibitemOpen
  \bibfield  {author} {\bibinfo {author} {\bibfnamefont {J.}~\bibnamefont
  {Lachaux}}, \bibinfo {author} {\bibfnamefont {E.}~\bibnamefont {Rodriguez}},
  \bibinfo {author} {\bibfnamefont {J.}~\bibnamefont {Martinerie}}, \bibinfo
  {author} {\bibfnamefont {F.}~\bibnamefont {Varela}},  \emph {et~al.},\
  }\href@noop {} {\bibfield  {journal} {\bibinfo  {journal} {Human brain
  mapping}\ }\textbf {\bibinfo {volume} {8}},\ \bibinfo {pages} {194} (\bibinfo
  {year} {1999})}\BibitemShut {NoStop}%
\bibitem [{\citenamefont {Arecchi}(2010)}]{arecchi2010}%
  \BibitemOpen
  \bibfield  {author} {\bibinfo {author} {\bibfnamefont {F.}~\bibnamefont
  {Arecchi}},\ }\href@noop {} {\bibfield  {journal} {\bibinfo  {journal}
  {Journal of Psychophysiology}\ }\textbf {\bibinfo {volume} {24}},\ \bibinfo
  {pages} {141} (\bibinfo {year} {2010})}\BibitemShut {NoStop}%
\bibitem [{\citenamefont {Rodriguez}\ \emph {et~al.}(1999)\citenamefont
  {Rodriguez}, \citenamefont {George}, \citenamefont {Lachaux}, \citenamefont
  {Martinerie}, \citenamefont {Renault},\ and\ \citenamefont
  {Varela}}]{rodriguez1999perception}%
  \BibitemOpen
  \bibfield  {author} {\bibinfo {author} {\bibfnamefont {E.}~\bibnamefont
  {Rodriguez}}, \bibinfo {author} {\bibfnamefont {N.}~\bibnamefont {George}},
  \bibinfo {author} {\bibfnamefont {J.-P.}\ \bibnamefont {Lachaux}}, \bibinfo
  {author} {\bibfnamefont {J.}~\bibnamefont {Martinerie}}, \bibinfo {author}
  {\bibfnamefont {B.}~\bibnamefont {Renault}}, \ and\ \bibinfo {author}
  {\bibfnamefont {F.~J.}\ \bibnamefont {Varela}},\ }\href@noop {} {\bibfield
  {journal} {\bibinfo  {journal} {Nature}\ }\textbf {\bibinfo {volume} {397}},\
  \bibinfo {pages} {430} (\bibinfo {year} {1999})}\BibitemShut {NoStop}%
\bibitem [{\citenamefont {K{\"o}rding}\ and\ \citenamefont
  {Wolpert}(2006)}]{kording2006}%
  \BibitemOpen
  \bibfield  {author} {\bibinfo {author} {\bibfnamefont {K.}~\bibnamefont
  {K{\"o}rding}}\ and\ \bibinfo {author} {\bibfnamefont {D.}~\bibnamefont
  {Wolpert}},\ }\href@noop {} {\bibfield  {journal} {\bibinfo  {journal}
  {Trends in cognitive sciences}\ } (\bibinfo {year} {2006})}\BibitemShut
  {NoStop}%
\bibitem [{\citenamefont {Poppel}(2004)}]{poppel2004lost}%
  \BibitemOpen
  \bibfield  {author} {\bibinfo {author} {\bibfnamefont {E.}~\bibnamefont
  {Poppel}},\ }\href@noop {} {\bibfield  {journal} {\bibinfo  {journal} {Acta
  Neurobiologiae Experimentalis}\ }\textbf {\bibinfo {volume} {64}},\ \bibinfo
  {pages} {295} (\bibinfo {year} {2004})}\BibitemShut {NoStop}%
\bibitem [{\citenamefont {Arecchi}(2011)}]{arecchi2011phenomenology}%
  \BibitemOpen
  \bibfield  {author} {\bibinfo {author} {\bibfnamefont {F.}~\bibnamefont
  {Arecchi}},\ }\href@noop {} {\bibfield  {journal} {\bibinfo  {journal}
  {Nonlinear Dynamics, Psychology and Life Sciences}\ }\textbf {\bibinfo
  {volume} {15}},\ \bibinfo {pages} {359} (\bibinfo {year} {2011})}\BibitemShut
  {NoStop}%
\bibitem [{\citenamefont {Koch}(2004)}]{koch2004}%
  \BibitemOpen
  \bibfield  {author} {\bibinfo {author} {\bibfnamefont {C.}~\bibnamefont
  {Koch}},\ }\href@noop {} {\emph {\bibinfo {title} {The quest for
  consciousness: A neuroscientific approach}}}\ (\bibinfo  {publisher} {Roberts
  \& Co},\ \bibinfo {year} {2004})\BibitemShut {NoStop}%
\bibitem [{\citenamefont {Dehaene}\ \emph {et~al.}(2003)\citenamefont
  {Dehaene}, \citenamefont {Sergent},\ and\ \citenamefont
  {Changeux}}]{dehaene2003}%
  \BibitemOpen
  \bibfield  {author} {\bibinfo {author} {\bibfnamefont {S.}~\bibnamefont
  {Dehaene}}, \bibinfo {author} {\bibfnamefont {C.}~\bibnamefont {Sergent}}, \
  and\ \bibinfo {author} {\bibfnamefont {J.}~\bibnamefont {Changeux}},\
  }\href@noop {} {\bibfield  {journal} {\bibinfo  {journal} {Proceedings of the
  National Academy of Sciences}\ }\textbf {\bibinfo {volume} {100}},\ \bibinfo
  {pages} {8520} (\bibinfo {year} {2003})}\BibitemShut {NoStop}%
\bibitem [{\citenamefont {P\"{o}ppel}(1997)}]{poppel1997hierarchical}%
  \BibitemOpen
  \bibfield  {author} {\bibinfo {author} {\bibfnamefont {E.}~\bibnamefont
  {P\"{o}ppel}},\ }\href@noop {} {\bibfield  {journal} {\bibinfo  {journal}
  {Trends in cognitive sciences}\ }\textbf {\bibinfo {volume} {1}},\ \bibinfo
  {pages} {56} (\bibinfo {year} {1997})}\BibitemShut {NoStop}%
\bibitem [{\citenamefont {P{\"o}ppel}(2009)}]{poppel2009pre}%
  \BibitemOpen
  \bibfield  {author} {\bibinfo {author} {\bibfnamefont {E.}~\bibnamefont
  {P{\"o}ppel}},\ }\href@noop {} {\bibfield  {journal} {\bibinfo  {journal}
  {Philosophical Transactions of the Royal Society B: Biological Sciences}\
  }\textbf {\bibinfo {volume} {364}},\ \bibinfo {pages} {1887} (\bibinfo {year}
  {2009})}\BibitemShut {NoStop}%
\bibitem [{\citenamefont {Brainard}(1997)}]{brainard1997}%
  \BibitemOpen
  \bibfield  {author} {\bibinfo {author} {\bibfnamefont {D.}~\bibnamefont
  {Brainard}},\ }\href@noop {} {\bibfield  {journal} {\bibinfo  {journal}
  {Spatial vision}\ }\textbf {\bibinfo {volume} {10}},\ \bibinfo {pages} {433}
  (\bibinfo {year} {1997})}\BibitemShut {NoStop}%
\bibitem [{\citenamefont {Pelli}(1997)}]{pelli1997}%
  \BibitemOpen
  \bibfield  {author} {\bibinfo {author} {\bibfnamefont {D.}~\bibnamefont
  {Pelli}},\ }\href@noop {} {\bibfield  {journal} {\bibinfo  {journal} {Spatial
  vision}\ }\textbf {\bibinfo {volume} {10}},\ \bibinfo {pages} {437} (\bibinfo
  {year} {1997})}\BibitemShut {NoStop}%
\bibitem [{\citenamefont {Kleiner}\ \emph {et~al.}(2007)\citenamefont
  {Kleiner}, \citenamefont {Brainard}, \citenamefont {Pelli}, \citenamefont
  {Ingling}, \citenamefont {Murray},\ and\ \citenamefont
  {Broussard}}]{kleiner2007}%
  \BibitemOpen
  \bibfield  {author} {\bibinfo {author} {\bibfnamefont {M.}~\bibnamefont
  {Kleiner}}, \bibinfo {author} {\bibfnamefont {D.}~\bibnamefont {Brainard}},
  \bibinfo {author} {\bibfnamefont {D.}~\bibnamefont {Pelli}}, \bibinfo
  {author} {\bibfnamefont {A.}~\bibnamefont {Ingling}}, \bibinfo {author}
  {\bibfnamefont {R.}~\bibnamefont {Murray}}, \ and\ \bibinfo {author}
  {\bibfnamefont {C.}~\bibnamefont {Broussard}},\ }\href@noop {} {\bibfield
  {journal} {\bibinfo  {journal} {Perception}\ }\textbf {\bibinfo {volume}
  {36}},\ \bibinfo {pages} {1} (\bibinfo {year} {2007})}\BibitemShut {NoStop}%
\bibitem [{\citenamefont {Neri}\ \emph {et~al.}(1998)\citenamefont {Neri},
  \citenamefont {Morrone},\ and\ \citenamefont {Burr}}]{neri1998seeing}%
  \BibitemOpen
  \bibfield  {author} {\bibinfo {author} {\bibfnamefont {P.}~\bibnamefont
  {Neri}}, \bibinfo {author} {\bibfnamefont {M.}~\bibnamefont {Morrone}}, \
  and\ \bibinfo {author} {\bibfnamefont {D.}~\bibnamefont {Burr}},\ }\href@noop
  {} {\bibfield  {journal} {\bibinfo  {journal} {Nature}\ }\textbf {\bibinfo
  {volume} {395}},\ \bibinfo {pages} {894} (\bibinfo {year}
  {1998})}\BibitemShut {NoStop}%
\bibitem [{\citenamefont {Wyszecki}\ and\ \citenamefont
  {Stiles}(1982)}]{wyszecki1982color}%
  \BibitemOpen
  \bibfield  {author} {\bibinfo {author} {\bibfnamefont {G.}~\bibnamefont
  {Wyszecki}}\ and\ \bibinfo {author} {\bibfnamefont {W.}~\bibnamefont
  {Stiles}},\ }\href@noop {} {\emph {\bibinfo {title} {Color science}}}\
  (\bibinfo  {publisher} {John Wiley \& Sons, New York},\ \bibinfo {year}
  {1982})\BibitemShut {NoStop}%
\bibitem [{\citenamefont {Arrighi}\ \emph {et~al.}(2009)\citenamefont
  {Arrighi}, \citenamefont {Arecchi}, \citenamefont {Farini},\ and\
  \citenamefont {Gheri}}]{farini16}%
  \BibitemOpen
  \bibfield  {author} {\bibinfo {author} {\bibfnamefont {R.}~\bibnamefont
  {Arrighi}}, \bibinfo {author} {\bibfnamefont {F.}~\bibnamefont {Arecchi}},
  \bibinfo {author} {\bibfnamefont {A.}~\bibnamefont {Farini}}, \ and\ \bibinfo
  {author} {\bibfnamefont {C.}~\bibnamefont {Gheri}},\ }\href {\doibase
  {10.1007/s10339-008-0244-9}} {\bibfield  {journal} {\bibinfo  {journal}
  {Cognitive Processing}\ }\textbf {\bibinfo {volume} {10}},\ \bibinfo {pages}
  {S95} (\bibinfo {year} {2009})}\BibitemShut {NoStop}%
\bibitem [{\citenamefont {Leopold}\ \emph {et~al.}(2002)\citenamefont
  {Leopold}, \citenamefont {Wilke}, \citenamefont {Maier},\ and\ \citenamefont
  {Logothetis}}]{Leopold:2002stable}%
  \BibitemOpen
  \bibfield  {author} {\bibinfo {author} {\bibfnamefont {D.~A.}\ \bibnamefont
  {Leopold}}, \bibinfo {author} {\bibfnamefont {M.}~\bibnamefont {Wilke}},
  \bibinfo {author} {\bibfnamefont {A.}~\bibnamefont {Maier}}, \ and\ \bibinfo
  {author} {\bibfnamefont {N.~K.}\ \bibnamefont {Logothetis}},\ }\href
  {\doibase 10.1038/nn851} {\bibfield  {journal} {\bibinfo  {journal} {Nat
  Neurosci}\ }\textbf {\bibinfo {volume} {5}},\ \bibinfo {pages} {605}
  (\bibinfo {year} {2002})}\BibitemShut {NoStop}%
\bibitem [{\citenamefont {Leopold}\ and\ \citenamefont
  {Logothetis}(1999)}]{leopold1999}%
  \BibitemOpen
  \bibfield  {author} {\bibinfo {author} {\bibfnamefont {D.}~\bibnamefont
  {Leopold}}\ and\ \bibinfo {author} {\bibfnamefont {N.}~\bibnamefont
  {Logothetis}},\ }\href@noop {} {\bibfield  {journal} {\bibinfo  {journal}
  {Trends in cognitive sciences}\ }\textbf {\bibinfo {volume} {3}},\ \bibinfo
  {pages} {254} (\bibinfo {year} {1999})}\BibitemShut {NoStop}%
\bibitem [{\citenamefont {Atmanspacher}\ \emph {et~al.}(2004)\citenamefont
  {Atmanspacher}, \citenamefont {Filk},\ and\ \citenamefont
  {R{\"o}mer}}]{atmanspacher2004}%
  \BibitemOpen
  \bibfield  {author} {\bibinfo {author} {\bibfnamefont {H.}~\bibnamefont
  {Atmanspacher}}, \bibinfo {author} {\bibfnamefont {T.}~\bibnamefont {Filk}},
  \ and\ \bibinfo {author} {\bibfnamefont {H.}~\bibnamefont {R{\"o}mer}},\
  }\href@noop {} {\bibfield  {journal} {\bibinfo  {journal} {Biological
  Cybernetics}\ }\textbf {\bibinfo {volume} {90}},\ \bibinfo {pages} {33}
  (\bibinfo {year} {2004})}\BibitemShut {NoStop}%
\end{thebibliography}%

\end{document}